\documentclass[nofootinbib,preprint]{revtex4}

\usepackage{graphicx}
\usepackage{amssymb}
\usepackage{amsmath}
\usepackage{bm}
\usepackage{hyperref}

\begin{document}

\title{Growth rate in inhomogeneous interacting vacuum models}
\author{Humberto A.~Borges, Patr\'icia Hepp}

\affiliation{Instituto de F\'isica, Universidade Federal da Bahia, Salvador, BA, 40210-340, Brasil}

\date{\today}

\begin{abstract}
We study the effects of vacuum energy perturbations on the evolution of the dark matter growth rate in a decomposed Chaplygin gas model with interacting dark matter and vacuum energy. We consider two different cases: (i) geodesic dark matter with homogeneous vacuum, and (ii) a covariant ansatz for vacuum density perturbations. In the latter case, we show that the vacuum perturbations are very tiny as compared to dark matter perturbations on sub-horizon scales. In spite of that, depending on the value of the Chaplygin gas parameter $\alpha$, vacuum perturbations suppress or enhance the dark matter growth rate as compared to the geodesic model.
\end{abstract}

\maketitle

\section{Introduction}\label{intro}

The recent accelerated expansion of the universe detected by the precision measurements of type Ia supernovae \cite{Astier,riess,perl}, anisotropies in the cosmic microwave background radiation \cite{spergel, g, ade, ade1} and baryon acoustic oscillations \cite{telubac, ata} indicates that about $96\%$ of the total energy of our universe is in the form of unknown dark fluid components, namely, dark energy \cite{tegmark} and dark matter, and the remaining components are in the form of baryonic matter and radiation. The clustering dark matter with zero pressure (cold dark matter) is concentrated in the local structures and plays crucial role for forming galaxies and clusters of galaxies while dark energy possesses a negative pressure that drives the recent accelerated expansion. The simplest model to describe dark energy is to associate it with a constant vacuum energy density $\rho_V$ characterised by the equation of state parameter $w=-1$, equivalent to a cosmological constant in Einstein gravity \cite{pebles,Padmanaban,weinberg02}. The cosmological model that incorporates a constant vacuum energy plus cold dark matter is known as $\Lambda CDM$. Despite its relative success when tested against the most precise observations, one of its problems is the large discrepancy between the current vacuum energy density, $\rho_V\sim 10^{-29} g/cm^3$, and the theoretical value predicted by quantum field theories \cite{weinberg02}. 

Among of many alternatives to circumvent the problem of the cosmological constant we can consider the generalised Chaplygin gas (gCg), a unified dark sector whose equation of state is given by \cite{kam, Fabris, Bento, Sandvik:2002jz, bento} 
\begin{equation}\label{g}
p=-\frac{A}{\rho^{\alpha}},
\end{equation}
where A is a constant, $\rho$ is the energy density and $\alpha$ is a free parameter.  This gCg interpolates between a cold dark matter dominating at early times and a dark energy component dominating at late times. However, due to a non-zero adiabatic sound speed, the perturbations exhibit strong instabilities and oscillations affecting radically the matter power spectrum, unless the parameter $\alpha$ does not differ too much from zero. An alternative to avoid this problem is to include a non-adiabatic pressure contribution in order to make the effective sound speed vanish \cite{amendola}. Another possibility is to split the gCg fluid into two interacting components, a pressuless cold dark matter with energy density $\rho_m$ and a vacuum term with equation of state $p_V = - \rho_V$ \cite{Zi, Wa, Ba}. In this way, the sound speed comes only from the vacuum energy perturbations, that need to be zero or negligible as compared to matter perturbations.

The above condition is reached if we assume that dark matter follows geodesics, with energy transfer proportional to its $4$-velocity, which implies that vacuum is homogeneous in the comoving-synchronous gauge \cite{HW, DW}. However, due to the dynamical nature of vacuum, it is important to consider the possibility of inhomogeneities in its energy density and verify explicitly if their fluctuations are in fact negligible at sub-horizon scales. This was done in a model equivalent to the generalised Chaplygin gas with $\alpha=$-1/2 \cite{W}, but here we generalise the analysis for any value of $\alpha$.

In this work we are interested to explore the impact of vacuum energy fluctuations on the evolution of the dark matter growth rate in a decomposed gCg model with interacting dark matter and vacuum energy. For the sake of simplicity, we neglect the contributions of the baryonic and radiation components. Under the assumption of a covariant ansatz for the vacuum energy density variation, we are able to compute the energy-momentum transfer up to first order using a gauge invariant perturbative approach. A scale-dependent second order differential equation for the dark matter density contrast is obtained, which allows us to follow the evolution of the linear dark matter growth rate, defined by
\begin{equation}\label{f}
f=\frac{\dot\delta_m}{H\delta\rho_m}.
\end{equation}

This paper is organised as follows. In section II we present the background evolution of the decomposed gCg model. In section III we perform a perturbative analysis of the evolution of the dark matter growth rate sourced by vacuum energy perturbations. A comparison is then made with the case of a homogeneous vacuum energy. In section IV we present our conclusions. Throughout this work, we assume the dimensionless density parameter of dark matter as $\Omega_{m0}=0.3$. For the particular case $\alpha=-1/2$ we use the concordance value $\Omega_{m0}=0.45$ \cite{C}.

\section{Decomposed Chaplygin gas}

In this section we present the background evolution of the decomposed gCg model as our interacting model. For this aim, consider the Friedmann equations in a spatially flat FLRW space-time,
\begin{eqnarray}
3H^2 = \rho, \label{Friedmann}\\
\dot\rho + 3H(\rho + p) = 0 \label{continuity},
\end{eqnarray}
where we are fixing $8\pi G= c=1$, $H$ is the Hubble parameter and a dot means derivative with respect to the cosmological time. 
Using the pressure $(\ref{g})$ into the conservation equation $(\ref{continuity})$ we can obtain the integrated solution for the energy density of the fluid,
\begin{equation}\label{total}
\rho=\bigg[A+\frac{B}{a^{3(1+\alpha)}}\bigg]^{-(1+\alpha)},
\end{equation} 
where $B$ is a constant of integration. Here, the present value $a_0$ of the scale factor was taken to be unity. This solution interpolates a matter dominated universe with standard solution $\rho\propto a^{-3}$ $(a\ll 1)$, which allows for structure formation, and a cosmological constant $\rho = A^{-(1+\alpha)}$ in the late epoch $(a\gg 1)$, leading to the cosmic acceleration. So the gCg model can be thought as a mixed of both dark energy and dark matter. For $\alpha=0$ the solution is the same as the $\Lambda$CDM model.

Let us split this fluid into two components, namely, cold dark matter with pressure $p_m=0$ and a vacuum-type term with equation of state $p_{V}=-\rho_V$ .
Hence, the decomposition implies that
\begin{eqnarray}\label{decomp}
\rho & = & \rho_m + \rho_V, \\
\rho_V& = & \frac{A}{\rho^\alpha}\label{decomp3}.
\end{eqnarray}
With the help of equations $(\ref{Friedmann})$, $(\ref{g})$ and $(\ref{decomp3})$ we are able to obtain the vacuum energy density 
\begin{equation}\label{pl}
\rho_{V}=\rho_{V 0}\bigg(\frac{H}{H_0}\bigg)^{-2\alpha}.
\end{equation}
It is straighforward to obtain the Hubble parameter and matter density, respectively
\begin{eqnarray}
H = H_0\bigg[1-\Omega_{m0}+\frac{\Omega_{m0}}{a^{3(1+\alpha)}}\bigg]^{\frac{1}{2(1+\alpha)}}, \label{Hubble}\\
\rho_{m}=3H^2-3H_{0}^{2(1+\alpha)}(1-\Omega_{m0})H^{-2\alpha}\label{Matter},
\end{eqnarray}
with
\begin{eqnarray}
A=\rho_{V 0}(3H_0^2)^{\alpha}, \label{deu}\\
B=(3H_0^2)^{(1+\alpha)}\left( 1-\frac{\rho_{V 0}}{3H_0^2} \right) \label{foila},
\end{eqnarray}
\begin{equation}\label{mms}
\Omega_{V 0}=\frac{\rho_{V 0}}{3H_{0}^2}, \ \ \  \Omega_{m 0}=\frac{\rho_{m0}}{3H_{0}^2},
\end{equation}
where a subindex 0 indicates the present value of the corresponding quantities. 

It is possible to put the conservation equation (\ref{continuity}) in the form
\begin{eqnarray}\label{continuity3}
\dot\rho_m + 3H\rho_m =Q, \\
\dot\rho_{V}=-Q,
\end{eqnarray}
where the energy transfer between the components is given by
\begin{equation}\label{Gamma}
Q = 6\alpha H_0(1-\Omega_{m 0})\bigg(\frac{H}{H_0}\bigg)^{-(2\alpha+1)}\dot{H}.
\end{equation}
The sign of $Q$ depends on the sign of the gCg parameter $\alpha$, since $\dot H<0$. If $Q>0$ the vacuum energy decays into cold dark matter, and in the opposite case cold dark matter decays into vacuum energy. Therefore, we see that, for $\alpha < 0$, the vacuum energy density decays along the expansion, while dark matter is created in the process, whereas for $\alpha > 0$ dark matter is annihilated. An interesting particular case is given by $\alpha = -1/2$. In this case,  the vacuum energy density decays linearly with $H$, while dark matter is created at a constant rate. On the other hand, for $\alpha = 0$ we re-obtain the standard model with a cosmological constant and conserved matter.

\section{The source term} 

We can explicitly write the source term $Q$ in a covariant manner for a perfect fluid described by two interacting dark components. For this purpose, let us assume a covariant form for the vacuum energy density $(\ref{pl})$, 
\begin{equation}\label{pls}
\rho_{V}=\rho_{V 0}\bigg(\frac{\Theta}{3H_0}\bigg)^{-2\alpha},
\end{equation}
where we use the scalar expansion $\Theta=u^{\mu}_{;\mu}$ with $u^{\mu}$ being the four velocity of the fluid. In the background universe the scalar expansion is $\Theta = 3H$.

The energy-momentum conservation equations for each component are given by
\begin{equation}\label{asd}
{T^{\nu\mu}_m}_{;\mu}=Q^{\nu},
\end{equation}
\begin{equation}\label{asdw}
{T^{\nu\mu}_V}_{;\mu}=-Q^{\nu},
\end{equation}
where
\begin{equation}
 T^{\mu\nu}_A=\rho_Au^{\mu}u^{\nu}+p_Ah^{\mu\nu}
 \end{equation}
is the energy-momentum tensor of each component, $h^{\mu\nu}=g^{\mu\nu}+u^{\mu}u^{\nu}$ is the projector tensor and $Q^{\mu}$ is the energy-momentum transfer between dark matter and vacuum. The latter can be decomposed parallel and perpendicularly to the four velocity $u^{\mu}$ as
\begin{equation}\label{dec}
Q^{\mu}=u^{\mu}Q+\bar{Q}^{\mu},
\end{equation}
with $Q=-u_{\mu}Q^{\mu}$, $\bar{Q}^{\mu}=h^{\mu}_{\nu}Q^{\nu}$, $u_{\mu}\bar{Q}^{\mu}=0$ and $u_{\mu}u^{\mu}=-1$.

Projecting equations $(\ref{asd})$ and $(\ref{asdw})$ parallel and perpendicularly to $u^{\mu}$, we find the energy conservation equations
\begin{equation}\label{lopq}
\rho_{m,\mu}u^{\mu}+\Theta\rho_m=-u_{\mu}Q^{\mu},
\end{equation}
\begin{equation}\label{lapis}
\rho_{V,\mu}u^{\mu}=u_{\mu}Q^{\mu},
\end{equation}
and the momentum conservation equations
\begin{equation}\label{hum}
\rho_m{u^{\mu}}_{;\nu}u^{\nu}=\bar{Q}^{\mu},
\end{equation}
\begin{equation}\label{huma}
\rho_{V,\nu}h^{\nu\mu}=-\bar{Q}^{\mu}.
\end{equation}
Using the ansatz $(\ref{pls})$ in the last equation, we find explicitly the covariant energy source term
\begin{equation}\label{werb}
Q=\frac{2}{3}\alpha(1-\Omega_{m 0})(3H_0)^{2(\alpha+1)}\Theta^{-(2\alpha+1)}{\Theta}_{,\mu}u^{\mu}.
\end{equation}
We complete our system of equations with the Raychaudhuri equation
\begin{equation}\label{sabe}
\Theta_{,\mu}u^{\mu}=-\frac{1}{3}\Theta^2-(u^{\mu}_{;\nu}u^{\nu})_{;{\mu}}+\frac{1}{2}(\rho_m-2\rho_V),
\end{equation}
where we have neglected the shear and vorticity contributions.
In the comoving frame, where the components of the four-velocity are $u_{0}=-1$, $u^{0}=1$ and $u_{i}=0=u^{i}$, one has $\bar{Q}^{\mu}=0$, which shows that there is no momentum transfer in the homogeneous and isotropic background. 

\subsection{Basic equations}

Now let us focus our attention to linear perturbations around a spatially flat FLRW universe. Let us start with the most general line element for scalar perturbations,
\begin{equation}
ds^2=-(1+2\phi)dt^2+2a^2B_{,i}dtdx^i+a^2[(1+2\psi)\delta_{ij}+2E_{,ij}dx^idx^j].
\end{equation}
The dark fluid velocity potencial $v$ can be defined by perturbing the four velocity $u_{\mu}=g_{\mu\nu}u^{\nu}$, which results in
\begin{equation}
\delta u_j=a^2\delta u^j+a^2B_j=v_{,j},
\end{equation}
assuming that $v$ is irrotational. We postulate that it coincides with the cold dark matter velocity potential $v_m$, since we cannot properly define the four velocity for the vacuum component.
The time component of the perturbed four-velocity is related to the perturbed metric through
\begin{equation}\label{estes}
\delta u_0=\delta u^0=-\phi.
\end{equation}

The next step is to obtain the conservation equations for each interacting dark component.
In order to provide a set of basic equations to calculate the matter density perturbation $\delta_m=\delta\rho_m/\rho_m$, we start by considering the equations for the vacuum.  The perturbation of the momentum equation $(\ref{huma})$ yields, in the comoving gauge, the result
\begin{equation}\label{momentum}
\partial^i\delta p_V^c = -\partial^i\delta \rho_V^c = -\delta\bar Q^i.
\end{equation}
So, a non-zero momentum transfer $\delta\bar Q^i$ is related to the presence of vacuum perturbations. Here the gauge invariant scalar quantities that characterise perturbations on comoving hypersurfaces were introduced,
\begin{equation}\label{mais}
\delta\mathcal A^c=\delta\mathcal A+\dot{\mathcal A}v.
\end{equation}
The perturbation of equation $(\ref{lapis})$ allows us to compute the energy transfer between the dark components, 
\begin{equation}\label{vai}
\delta Q^c= \dot\rho_V(\dot v+\phi) -\delta\dot\rho^c_V.
\end{equation}

For the dark matter component, the energy balance $(\ref{lopq})$ and the momentum balance $(\ref{hum})$  can be written, up to first order, respectively as
\begin{equation}\label{mn}
\dot\delta_m^c+\frac{Q}{\rho_m}\delta_m^c+\delta\Theta^c=\frac{\delta Q^c}{\rho_m}+\bigg(\frac{Q}{\rho_m}-3H\bigg)(\dot v+\phi),
\end{equation}
\begin{equation}\label{sei}
(\dot v+\phi)_{, j}=\frac{\delta\bar Q_{j}^c}{\rho_m}.
\end{equation}
The latter shows that, if the momentum transfer $\delta\bar Q_j^c$ is non zero, the dark matter particles are forced to deviate from their geodesic motions. This means that the evolution of the matter perturbation $\delta_m^c$ should be affected by the background evolution and the source terms owing to vacuum inhomogeneities.

To complete our system of equations, the Raychaudhuri equation for the expansion is obtained from the perturbation of $(\ref{sabe})$,
\begin{equation}\label{opai}
\delta\dot\Theta^c+\frac{2}{3}\Theta\delta\Theta^c+\frac{1}{2}\rho_m\delta_m^c=\delta\rho_V^c+\bigg(\frac{\nabla^2}{a^2}+\dot\Theta\bigg)(\dot v+\phi).
\end{equation}
For investigating the possibility of non-zero vacuum perturbations $\delta\rho^c_{V}$ and how they affect structure formation [7], we need now to specify a precise form for the energy and momentum transfer.

\subsection{Geodesic model}

Firstly, we assume that the energy transfer between the dark components follows the dark matter velocity, $Q^{\mu}=Qu^{\mu}$. In this case the momentum transfer $\bar Q^{\mu}$ is zero at the background and perturbative levels, which implies that the dynamical vacuum is homogeneous, with $\delta\rho^c_V=0$ according to $(\ref{momentum})$. Consequently, dark matter particles follow geodesics in a comoving frame. Furthermore, there is no energy transfer at first order, as we can see from $(\ref{sei})$ and $(\ref{vai})$.  So, the basic equations that describe the dynamics of the matter perturbations and scalar expansion are given by $(\ref{mn})$ and $(\ref{opai})$ in the absence of source terms,
\begin{equation}\label{eq1}
\dot\delta_m^c+\frac{Q}{\rho_m}\delta_m^c+\delta\Theta^c=0,
\end{equation}
\begin{equation}\label{opai1}
\delta\dot\Theta^c+\frac{2}{3}\Theta\delta\Theta^c+\frac{1}{2}\rho_m\delta_m^c=0.
\end{equation}
These equations are the same obtained in the synchronous comoving gauge [8], and a simpler second order differential equation for the density contrast can be found. To do that, we 
differentiate the continuity equation $(\ref{eq1})$ with respect to time and eliminate $\delta\Theta^c$ and $\delta\dot\Theta^c$ by using $(\ref{eq1})$ and $(\ref{opai1})$, to obtain 
\begin{equation}\label{sin}
\ddot\delta_m^c+\bigg[\frac{Q}{\rho_m}+2H\bigg]\dot\delta_m^c+\bigg[\frac{d}{dt}\bigg(\frac{Q}{\rho_m}\bigg)+2H\frac{Q}{\rho_m}-\frac{1}{2}\rho_m\bigg]\delta_m^c=0.
\end{equation}
The function $Q/\rho_m$ is the rate of homogeneous creation/annihilation of dark matter. For the standard $\Lambda$ CDM model ($\alpha =0$ ) this quantity is zero and dark matter is independently conserved.  For those interacting models in which $Q/\rho_m >0$, corresponding to values of $\alpha <0$, we have energy exchange from vacuum to dark matter and dark matter is created. In the case $Q/\rho_m <0$, corresponding to $\alpha >0$, we have energy flux from dark matter to vacuum and dark matter is annihilated.

\begin{figure}
\centerline{\includegraphics[height=4.5cm]{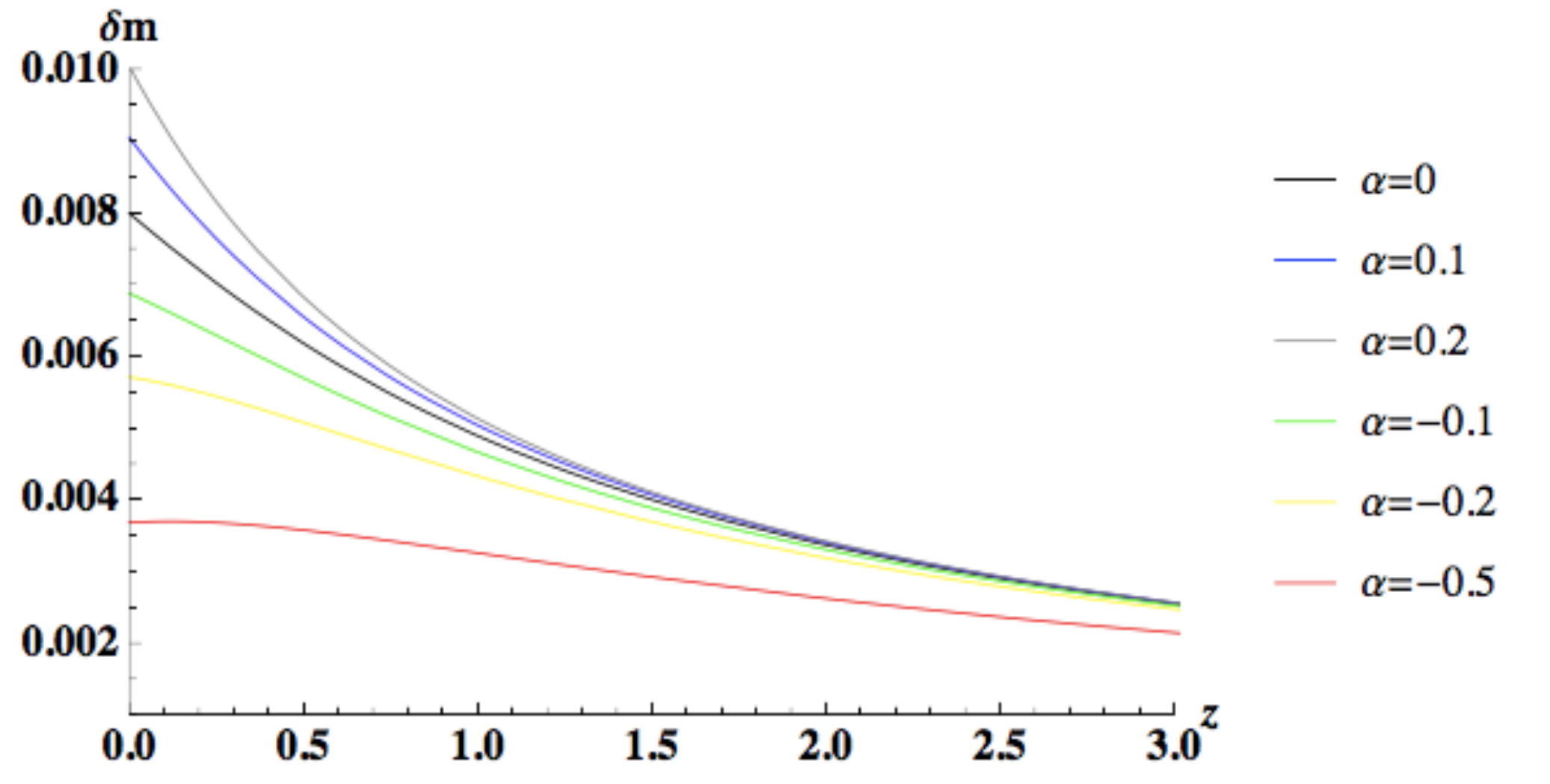} \hspace{.2in} \includegraphics[height=4.5cm]{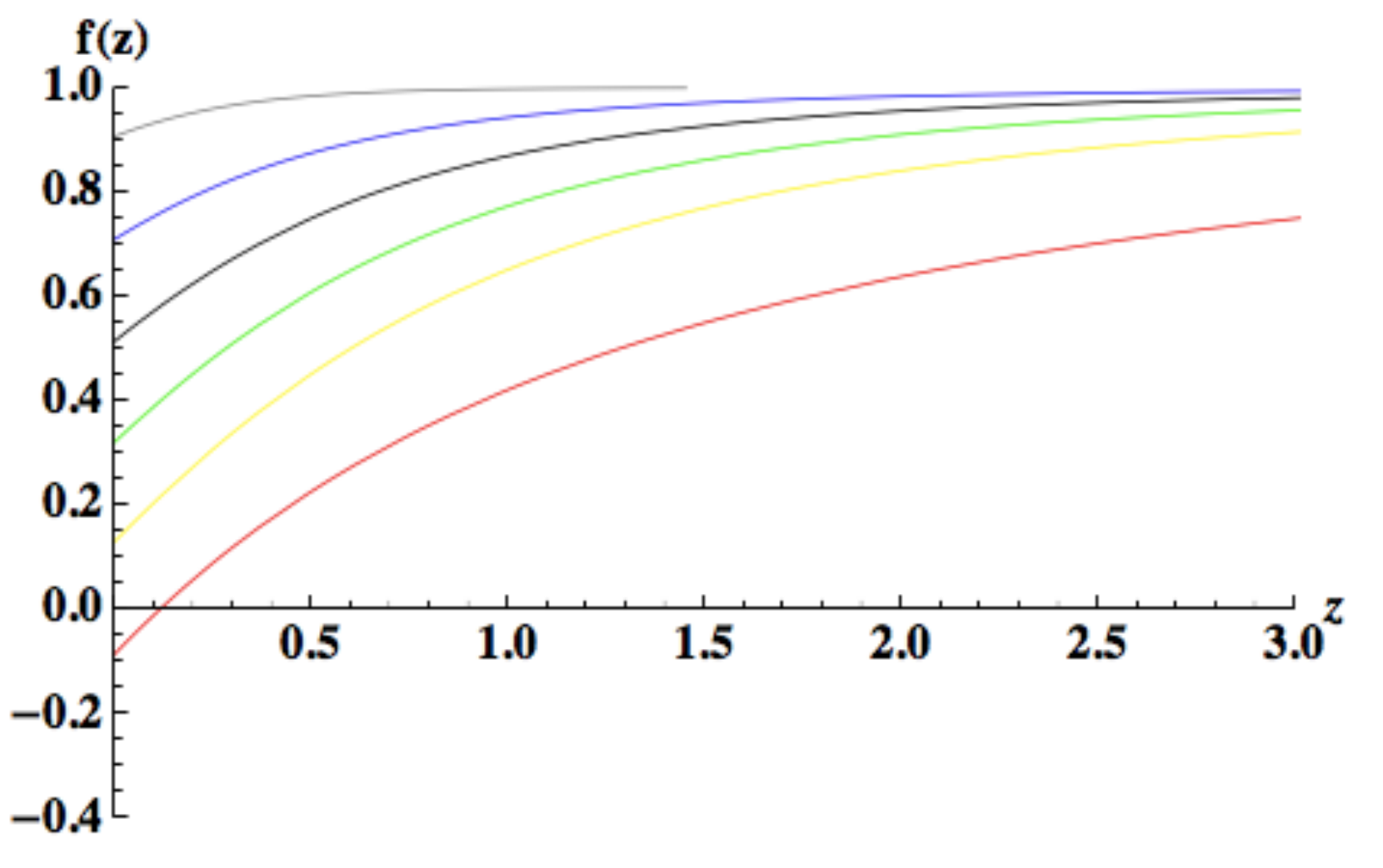}}
\caption{Dark matter density contrast and growth rate for the $\Lambda$CDM model (black) and for interacting models as indicated. We have used the best fit value $\Omega_m=0.45$ for the model with $\alpha=-0.5$. For all the other models we have used $\Omega_m=0.3$.}
\end{figure}

We are interested in computing the evolution of the linear growth rate $f$ defined by {(\ref{f})}, in order to illustrate how the creation/annihilation of dark matter can affect LSS data, for instance the redshift space distortions caused by the peculiar velocities of galaxies. These peculiar velocities distort the observed redshift maps, introducing an anisotropy in the clustering of galaxies in redshift space. Since the matter distribution determines the velocity, we can use this effect as a way to probe the linear growth rate $f$ and, in this way, differentiate between homogeneous and inhomogeneous vacuum energy.

To obtain the evolution of $f$ we need to solve the second order equation $(\ref{sin})$ by fixing the same initial amplitude $\delta_m(z_i)$ for all models at $z_i=1000$ noting that in the matter dominated era we have the standard value $f(z_i)=1$, since the dark matter contrast is proportional to the scale factor, $\delta_m \propto a$.

In Fig. 1 the evolution of both dark matter contrast and growth rate are depicted for the $\Lambda$CDM model ($\alpha = 0$) and for five interacting models corresponding to $\alpha = 0.1, 0.2, -0.1, -0.2, -0.5$. We see that, when compared with the $\Lambda$CDM model, both $\delta_m^c$ and $f$ are suppressed for models with $\alpha <0$ due to the homogeneous creation of dark matter, while are enhanced for the models with $\alpha >0$.

\subsection{Inhomogeneous vacuum model}

An alternative choice is to explicitly consider inhomogeneities in the vacuum, since neglecting them may lead to false interpretations of the observations \cite{Chan}. The natural manner for calculating the energy-momentum transfer at the perturbative level is to assume the covariant ansatz $({\ref{pls}})$ for the vacuum energy density, such that the perturbation of this quantity up to first order is related to the scalar expansion through the expression
\begin{equation}\label{sei23}
\delta\rho_V^c=\frac{2Q}{3\rho_m}\delta\Theta^c.
\end{equation}
This is our basic assumption. The relation above, together with the equation of state for vacuum, can be used into $(\ref{momentum})$ to obtain the right-hand side of the momentum equation $(\ref{sei})$, given by
\begin{equation}\label{paises4}
\delta\bar Q_j={\delta\rho_V^c}_{,j}.
\end{equation}
Perturbing $(\ref{werb})$ and using the Raychaudhuri equation $(\ref{opai})$ and the relations above, it is possible to write the energy transfer function in the Fourier space as
\begin{equation}\label{open45}
\frac{\delta Q^c}{\rho_m}=\frac{Q}{3\rho_m}\delta_m^c+\bigg[2H-\frac{2Q}{3\rho_m}+\frac{2QH^2}{3\rho_m^2}\bigg(\frac{k}{aH}\bigg)^2-\frac{(2\alpha+1)\rho_m}{2H}\bigg]\frac{\delta\rho_{\Lambda}^c}{\rho_m},
\end{equation}
where $k$ is the comoving wave number. The scale dependence that appears in the third term into the brackets is due to the momentum transfer between the dark components, owing to the presence of vacuum perturbations.
The amplitude of these perturbations compared to the dark matter perturbations can be evaluated by using $(\ref{mn})$ together with $(\ref{sei})$, $(\ref{paises4})$ and $(\ref{open45})$, leading to
\begin{equation}\label{fui}
\frac{\delta\rho_{\Lambda}^c}{\delta\rho_m^c}=-\frac{2Q}{3\rho_m^2 K}\bigg[Hf+\frac{2Q}{3\rho_m}\bigg].
\end{equation} 
Here we have defined the scale dependent function
\begin{equation}\label{poise}
K(a,k)=1-\frac{2Q}{3\rho_m^2}\bigg[A-H-\frac{(2\alpha+1)\rho_m}{2H}\bigg],
\end{equation}
where
\begin{equation}
A(a,k)=\frac{Q}{3\rho_m}+\frac{2QH^2}{3\rho_m ^2}\bigg(\frac{k}{aH}\bigg)^2 .
\end{equation}

The Raychaudhuri equation $(\ref{opai})$ can be written as
\begin{equation}\label{opai3}
{\delta\dot\Theta}^c=-\frac{1}{2}\rho_m\delta_m-2H\delta\Theta^c-\bigg[\frac{H^2}{\rho_m}\bigg(\frac{k}{aH}\bigg)^2+\frac{1}{2}\bigg]\delta\rho_{\Lambda}^c.
\end{equation}
Now we can differentiate $(\ref{fui})$, eliminate $\delta \dot{\Theta}^c$ through the perturbed Raychaudhuri equation $(\ref{opai3})$ and $\delta\Theta^c$ through $(\ref{fui})$ and $(\ref{sei23})$, to obtain a second order differential equation for the evolution of the dark matter contrast,
\begin{equation}\label{bo}
\ddot{\delta}_m^c+\bigg[\frac{2Q}{3\rho_m}+2H+\bigg(A-\frac{\dot K}{K}\bigg)\bigg]\dot{\delta}_m^c+\bigg[\frac{d}{dt}\bigg(\frac{2Q}{3\rho_m}\bigg)+2H\bigg(\frac{2Q}{3\rho_m}\bigg)-\frac{1}{2}\rho_m K+\frac{2Q}{3\rho_m}\bigg(A-\frac{\dot K}{K}\bigg)\bigg]\delta_m^c=0.
\end{equation}
The above equation is central result of the present paper. We see that differences arises as compared to the homogeneous vacuum model $(\ref{sin})$, namely, a reduction by a factor $2/3$ in the creation/annihilation rate and a change in the evolution of the dark matter contrast through the scale-dependent function $K$. Furthermore, a new function $A-\frac{\dot K}{K}$ appears in the coefficients of $\dot\delta_m$ and $\delta_m$. The standard $\Lambda$CDM model is recovered if we choose $\alpha=0$.

To estimate the importance of vacuum energy perturbations relative to dark matter as given by expression $(\ref{fui})$, we start by looking for their values in the deeper matter dominated phase ($z \gg 1$). Since $H\gg Q/\rho_m$ at high redshifts, from $(\ref{poise})$ we have $K \approx 1$, and hence the density contrast is proportional to the scale factor, $\delta_m\propto a$, resulting in the standard growth rate $f=1$. So, the expression $(\ref{fui})$ assumes the scale-independent form
\begin{equation}
\frac{\delta\rho_{\Lambda}^c}{\delta\rho_m^c}\approx \frac{2Q}{3\rho_m}\Omega_{m0}^{-\frac{1}{2(1+\alpha)}}z^{-3/2}.
\end{equation}
This ratio is very tiny and depends essentially on the interaction rate and the present value of the dark matter density. For comparison purposes, if we assume the model with $\alpha = -0.5$ and $\Omega_{m0}=0.45$, corresponding to a constant interaction rate, we found $\frac{\delta\rho_{\Lambda}^c}{\delta\rho_m^c}\sim 10^{-5}$ at $z_i=1000$. On the other hand, at the same redshift, for $\alpha = -0.1$ and $\Omega_{m0}=0.3$ we have $\frac{\delta\rho_{\Lambda}^c}{\delta\rho_m^c}\sim 10^{-9}$. 

\begin{figure}
\centerline{\includegraphics[height=5.5cm]{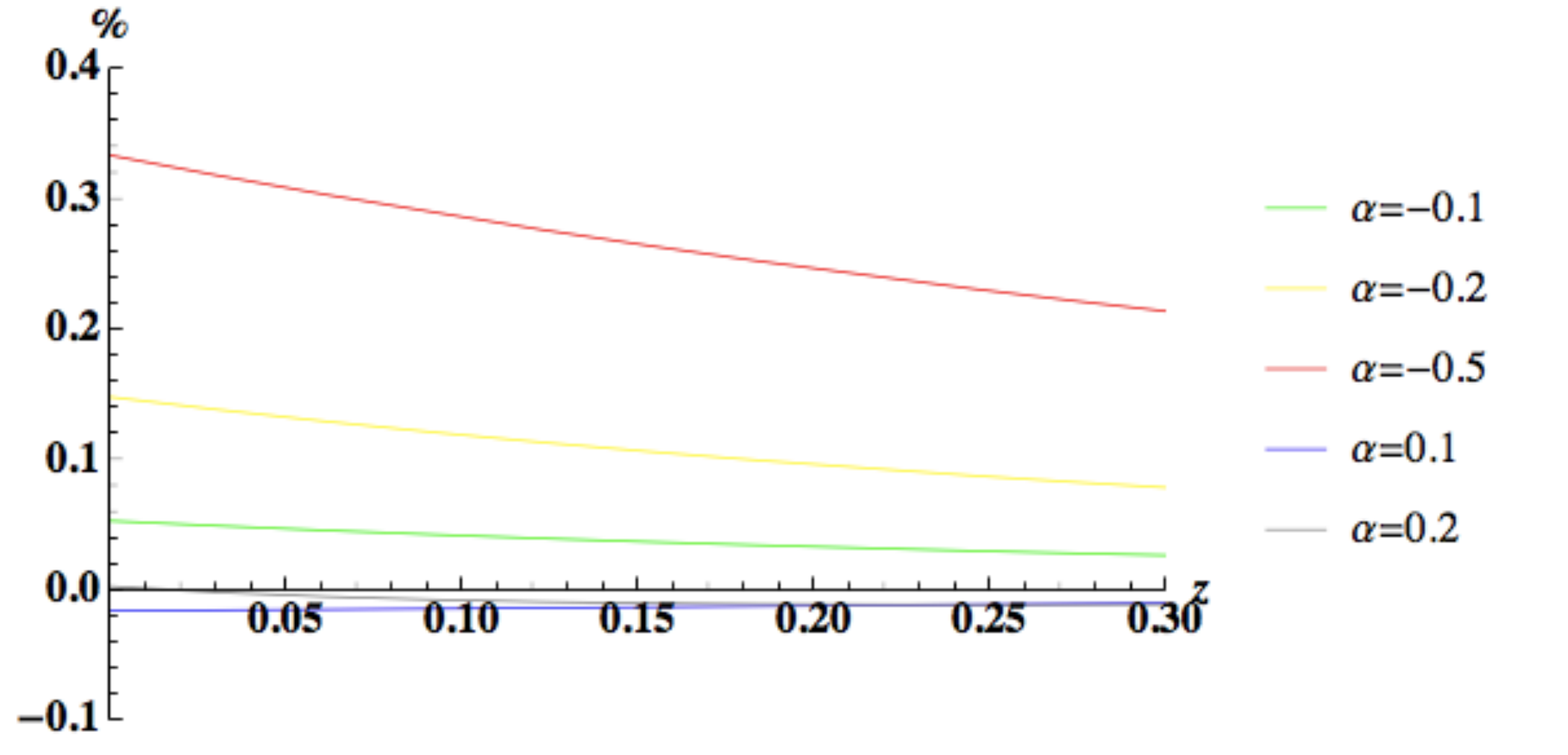}}
\caption{Relative difference between the dark matter contrasts for the scale $k$=0.2 and for the scale $k$=0.01, as a function of the redshift for different values of the gCg parameter.}
\end{figure}

\begin{figure}
\centerline{\includegraphics[height=5cm]{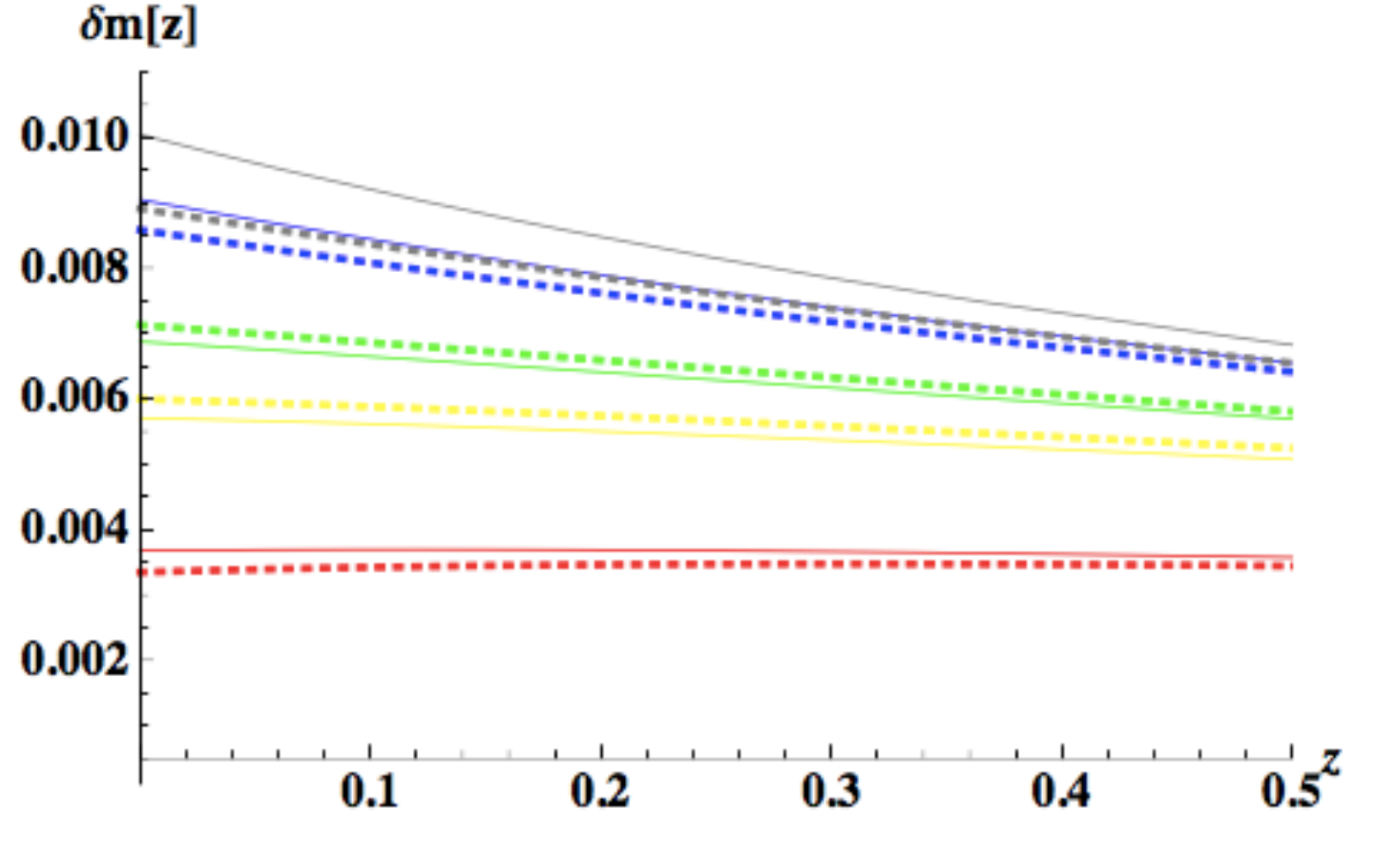} \hspace{.05in} \includegraphics[height=5cm]{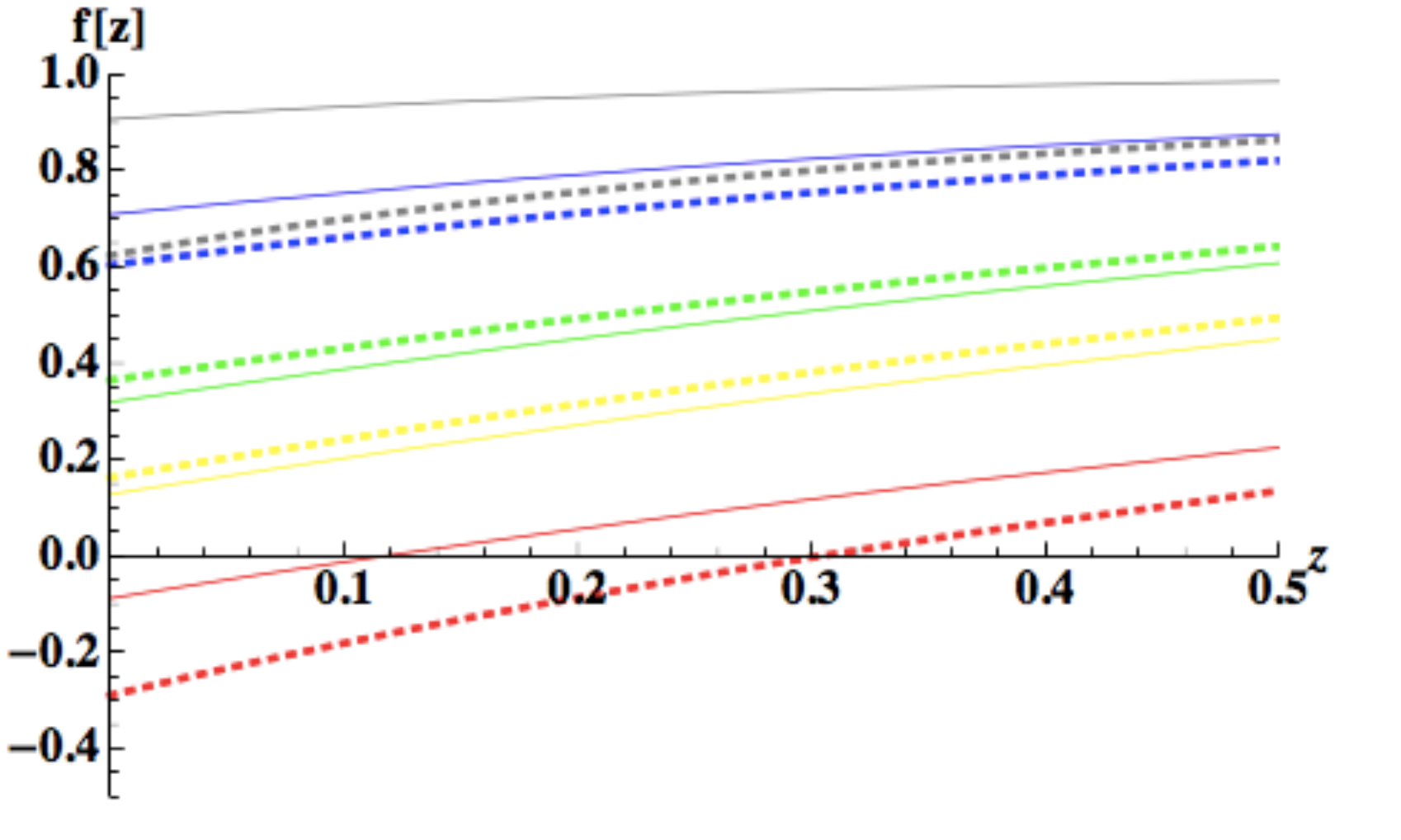}}
\caption{Dark matter contrast (left panel) and growth rate (right panel) for the following models: $\alpha =-0.5$ (red), $\alpha =-0.2$ (yellow), $\alpha =-0.1$ (green), $\alpha =0.1$ (blue) and $\alpha =0.2$ (gray). For all models we have used $\Omega_m=0.3$. Solid curves correspond to the geodesic model and dotted curves to inhomogeneous vacuum.}
\end{figure}

At late times the vacuum perturbations depend on the scale. The observational data of the linear power spectrum lie in the comoving wave number range $0.01$ Mpc$^{-1}<k<0.2$ Mpc$^{-1}$. In this range, taking the gCg parameter in the interval $-0.5 \leq\alpha\leq 0.5$, we find  the ratio between the vacuum and dark matter perturbations at $z=0$ in the interval
\begin{equation}
10^{-6} < \frac{\delta\rho_{\Lambda}^c}{\delta\rho_m^c} < 10^{-4},
\end{equation}
where the upper value corresponds to $k=0.01$, and the lower value to $k=0.2$. Therefore, vacuum perturbations are strongly suppressed inside the Hubble horizon respect to dark matter perturbations for the class of interacting models considered here. The smaller the scale, the stronger the suppression. 

These results are shown in Fig. 2, where we plot the ratio between dark matter perturbations for the scale $k$=0.2 and for the scale $k$=0.01 as a function of the redshift, which is less than $0.4\%$ for $|\alpha| < 0.5$. This allows us to state that vacuum perturbations are actually negligible on scales relevant for cosmic structure formation. Therefore, we can assume a perfectly homogeneous dynamical vacuum as a good approximation, such that dark matter follows geodesics. But we must verify the effects of vacuum perturbations on the evolution of the dark matter growth rate $f$ as compared to the homogeneous model.

 Fig. 3 shows the dark matter density contrast and growth rate when we use the geodesic model (solid curves) or the inhomogeneous vacuum model (dotted curves). The differences increase for large values of $|\alpha|$. We see that vacuum perturbations yield an enhancement in the curves as compared to the geodesic model for $\alpha = -0.1, -0.2$, and a suppression for $\alpha =0.1, 0.2$. For the case $\alpha = -0.5$, corresponding to dark matter creation at constant rate, a large suppression appears in the dark matter growth rate. Therefore, the correction terms introduced in equation $(\ref{bo})$ by the vacuum inhomogeneities should be taken into account when we analyse the growth rate evolution.

\section{Conclusions}

In this work we investigated the evolution of dark matter perturbations in the context of an interacting dark sector corresponding to a decomposed Chaplygin gas model. We have considered two distinct models for the covariant energy-momentum transfer. In the first, geodesic model the energy transfer follows the dark matter $4$-velocity. In this simplest case the momentum transfer is zero, which implies a homogeneous vacuum energy. We recover the scale-independent second order differential equation for the density contrast, showing that, compared to the $\Lambda$CDM cosmology, the growth rate is suppressed for $\alpha < 0$ due to the homogeneous creation of dark matter, and enhanced when $\alpha > 0$. 

In the second model, the momentum transfer is determined by the gradient of the vacuum perturbations, which is proportional to the scalar expansion, $\delta\rho^c_V\propto\delta\Theta^c$. The dynamics in this case is reduced to a single scale-dependent second order equation. We are able to evaluate the size of vacuum energy perturbations compared with the dark matter ones, determining the evolution of the growth rate for diverse gCg background solutions. The vacuum perturbations show to be negligible on scales inside the horizon, which implies that the density contrast of dark matter can be treated as scale-independent as is the case of the geodesic model. However, the vacuum perturbations affect the evolution of the growth rate as compared to the geodesic model, in a way that depends on the sign and value of $\alpha$. We then conclude that different perturbative models for the energy-momentum transfer may lead to different evolutions of the growth rate.

\section*{Acknowledgements}

The authors are thankful to Saulo Carneiro and Winfried Zimdahl for a critical reading.

\end{document}